\documentclass[12pt,prd,showpacs,amsmath,amssymb,aps,floats]{revtex4-1}%
\usepackage{epsfig}
\usepackage{dcolumn}
\usepackage{bm}
\usepackage{amsmath}
\usepackage{amsfonts}
\usepackage{amssymb}
\usepackage{longtable}
\def\nnd{\end{document}}

\def\be{\begin{equation}}
\def\ee{\end{equation}}

\newcommand{\bea}{\begin{eqnarray}}
\newcommand{\eea}{\end{eqnarray}}
\newcommand{\bwt}{\begin{widetext}}
\newcommand{\ewt}{\end{widetext}}

\def\u
\def\hZ{\widehat Z}

\def\eed{\end{document}}

\def\m_z{m_{\textrm {Z}}}

\renewcommand{\u}{\rm{u}}

\def\be{\beta}

\def\sl#1{#1\!\!\!\!/}
\def\gev{\textrm{GeV}}

\def\rm#1{\textrm{#1}}

\makeatother
\begin{document}
\title{Searches for the $t^\prime$ of a fourth family}
\author{Bob Holdom$^1$}
\author{Qi-Shu Yan$^2$}
\affiliation{$^1$ Department of Physics, University of Toronto, Toronto, Ontario M5S1A7, Canada }
\affiliation{$^2$ College of Physics Sciences, Graduate University of Chinese Academy of Sciences, \\ Beijing 100039, P.R China }

\begin{abstract}
We study the detection of the $t^\prime$ of a fourth family during the early running of LHC with 7 TeV collision energy and 1 fb$^{-1}$ integrated luminosity. By use of a neural network we show that it is feasible to search for the $t'$ even with a mass close to the unitarity upper bound, which is in the $500$ to $600$ GeV range. We also present results for the Tevatron with $10 \,\, \textrm{fb}^{-1}$. In both cases the search for a fourth family quark doublet can be significantly enhanced if one incorporates the contribution that the $b'$ can make to a $t'$-like signal. Thus the bound on the mass of a degenerate quark doublet should be stronger than the bounds obtained by treating $t'$ and $b'$ in isolation.
\end{abstract}

\maketitle

\renewcommand{\thefootnote}{\arabic{footnote}} \setcounter{footnote}{0}%

The LHC has begun its historic mission as it collects at least 1 fb$^{-1}$ of data at 7 TeV collision energy. For the sequential fourth family model \cite{Frampton:1999xi,Holdom:2009rf} the search for heavy quarks at the LHC will be the essential test, although other low energy experiments like LHCb \cite{Mohanta:2010eb} can help to bound the model parameters. The fourth family contains the doublet $(t',b')$, and when their masses are in the 500 to 600 GeV range they are close to an upper bound imposed by partial wave unitarity \cite{Chanowitz:1979}. Fourth family quark masses anywhere close to this range would have significant implications for electroweak symmetry breaking and flavor physics \cite{Holdom:2006mr}.

We shall assume that the dominant decay modes for these two heavy quarks are $t^\prime \to W b$ and $b^\prime \to t W \to W W b$, which is consistent with current bounds \cite{Soni:2010xh,Chanowitz:2010bm}. It is known that the multivariate analysis methods \cite{Hocker:2007ht}, like the neural network method and the boost decision tree method, can be quite useful to separate signal to background. They have been used successfully in the top quark precision measurements \cite{Pleier:2008ig} and for single top production at Tevatron \cite{Abazov:2009ii,Aaltonen:2009jj}. In our analysis we use the MLP (Multi-Layer Perceptron) neural network method available in the TMVA package \cite{Hocker:2007ht}.

A feature of $t'$ decay is the production of highly boosted and isolated $W$'s. The hadronic decay of such jets produce $W$-jets, and their use in a simple reconstruction of the $t'$ was explored in \cite{Holdom:2007nw,Holdom:2007ap,Holdom:2010}. (The $W$-jets can also be searched for on their own \cite{Holdom:2010}.) Since these results were encouraging we were prompted to consider how a more conventional full reconstruction method could be enhanced with a neural network. Recently a similar approach was used for the more difficult process $p p\to b^\prime {\bar b^\prime} \to \ell \nu 8 j$ \cite{Holdom:2010fr} at $10$ TeV and $1$ fb$^{-1}$ with $m_{b^\prime}=600$ GeV. The sensitivity was found to compare favorably with  the same-sign lepton mode. 

The CDF collaboration with $4.6$ fb$^{-1}$ of data has produced the bound $m_{t^\prime} > 335 \textrm{GeV}$ by using a two dimensional fit to the ($M_{rec}$, $H_T$) distribution \cite{cdfnote}. An early investigation \cite{atlas} of a $t^\prime$ search at LHC 
would have very pessimistic implications for the success of a $t'$ search in the unitarity region in the first early running data. A more recent study of a vector-like quark decay $T\rightarrow Wb$ in \cite{AguilarSaavedra:2009es}, also carried out at 14 TeV, has less pessimistic conclusions. In that study two tagged $b$ jets were required and a likelihood discrimination analysis was adopted. 

In this work we revisit the sensitivity to $t'$ during the early running of LHC by proposing a new reconstruction method in association with a neural network. We account for the effect that $b^\prime {\bar b^\prime}$ production can have on a $t'$ search, which has not been done elsewhere. Depending on how the signal is defined, the $b'$ can strengthen  a signal above standard model backgrounds. In our analysis we assume degenerate $t'$ and $b'$ masses. The impact of the $b'$ on the $t'$ analysis will only grow if the former has a smaller mass, as is often assumed, and thus we are providing a conservative estimate of this impact. Degenerate quark masses have not yet been ruled out in a model independent analysis. For example it is often assumed that the fourth neutrino has a Dirac mass, but when it has a Majorana mass then the constraints on mass ratios change considerably \cite{Holdom:2006mr,Holdom:2010}. Also, for higher quark masses a simple perturbative analysis need no longer apply.

We use Madgraph/MadEvent \cite{Maltoni:2002qb} to generate signal events and Alpgen \cite{Mangano:2002ea} to generate background events. The MLM parton-jet matching method is used with $p_{T\rm{min}}=100$ GeV for the $t {\bar t}+ nj$ samples and $p_{T\rm{min}}=150$ GeV for the $W + nj$ samples. These choices are reasonable given the high value of $H_T\sim2m_{q'}$. In principle there should be little dependence on $p_{T\rm{min}}$ and we shall test this further below. Pythia \cite{Sjostrand:2006za} is used to simulate shower, fragmentation, hadronization and decay processes. PGS \cite{pgs} is used to simulate the detector effects and to find jets, leptons, and missing energy in each event. We modify the PGS code slightly to use the anti-$k_T$ jet-finding algorithm \cite{Cacciari:2008gp}. For the jet resolution parameter we choose $R=0.4$. Other possible backgrounds, and in particular the irreducible $t\overline{b}W$ background, have been found to be small \cite{Holdom:2007ap}.

We adopt a few preselection rules: jets are required to have $p_T(j)>20$ GeV, there is only one energetic lepton with $p_T(\ell)>20$ GeV ($\ell=e\,,\,\,\mu$) and the missing energy satisfies $\sl{E}>20$ GeV. In Table \ref{table4} we provide the selection efficiency of the preselection cuts. After the preselection cuts the backgrounds are about an order of magnitude larger than the signal. Note that in order to generate the relevant background events more efficiently we have imposed process dependent $H_T$ cuts, chosen in such a way as not to significantly affect the final results. 
\begin{table}[th]
\begin{center}%
\begin{tabular}
[c]{|c|c|c|c|c|}\hline
& $b^\prime \bar{b^\prime}$ & $t^\prime \bar{t^\prime}$& $W + \textrm{jets}$ &  $t {\bar t} + \textrm{jets}$ \\\hline
$p_T(\ell)>20\rm{GeV}$ & $37\%$ & $29\%$ & $50\%$ & 
$26\%$  \\ \hline
$\sl{E}>20\rm{GeV}$ & $36\% $ & $28\%$ & $33\%$ &
$24\%$  \\ \hline
\hline
Events with 1 fb$^{-1}$ & $112.8$ & $85.6$ & $881.8$ & $1285.3$ \\\hline
\end{tabular}
\end{center}
\caption{Selection efficiencies of preselection rules are demonstrated, where in the last row we normalize the number of survived events for 1 fb$^{-1}$ while assuming $n_j \geq 3$. We also present the percentage for each rules in the preselection. In this table, the $K$ factors, 1 for $W+$ jets and 1.5 for the rest, have been included.}%
\label{table4}%
\end{table}

\begin{figure}[h]
\centerline{
\epsfxsize=7 cm \epsfysize=5 cm \epsfbox{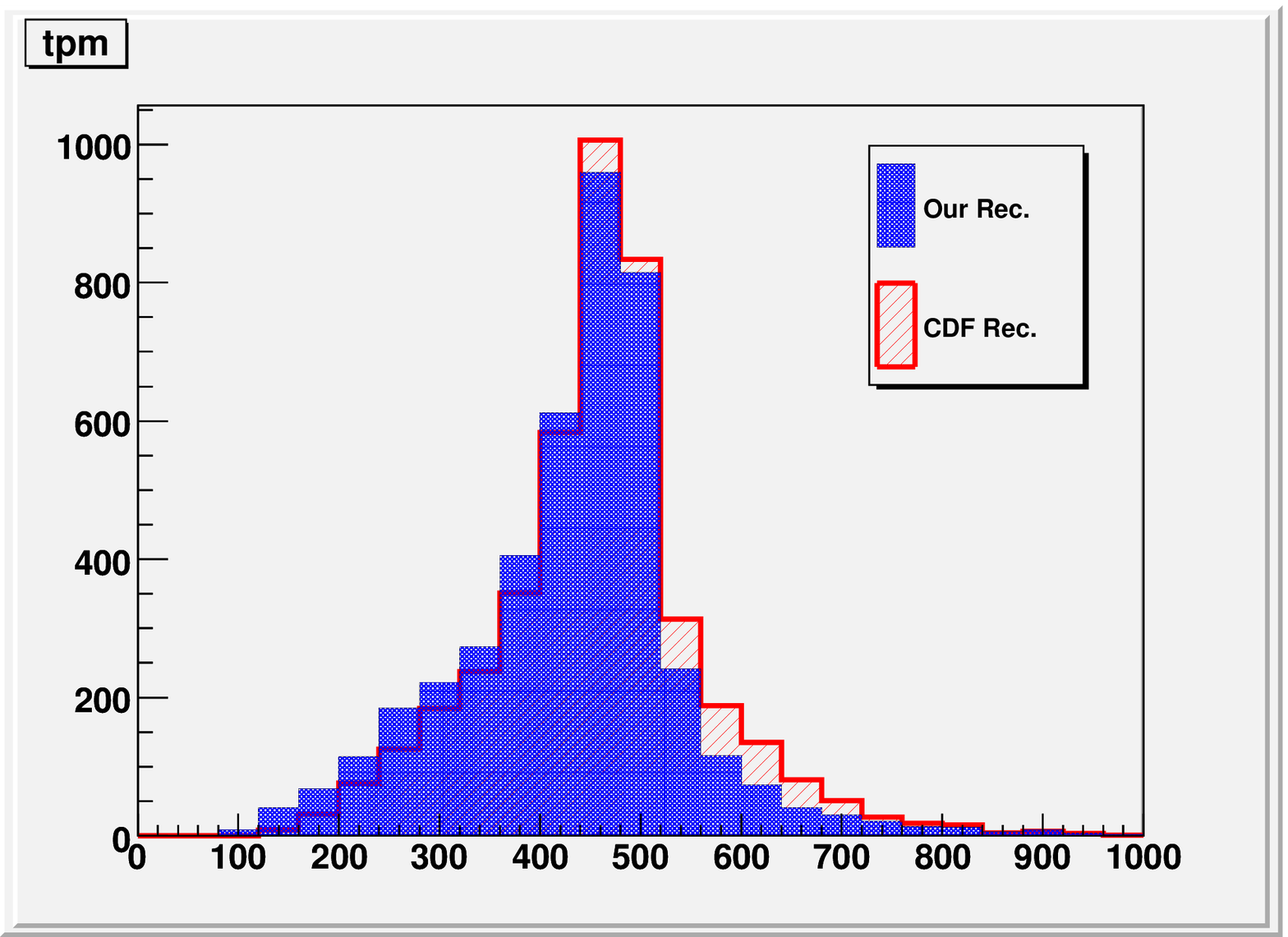}
\hspace*{0.2cm}
\epsfxsize=7 cm \epsfysize=5 cm \epsfbox{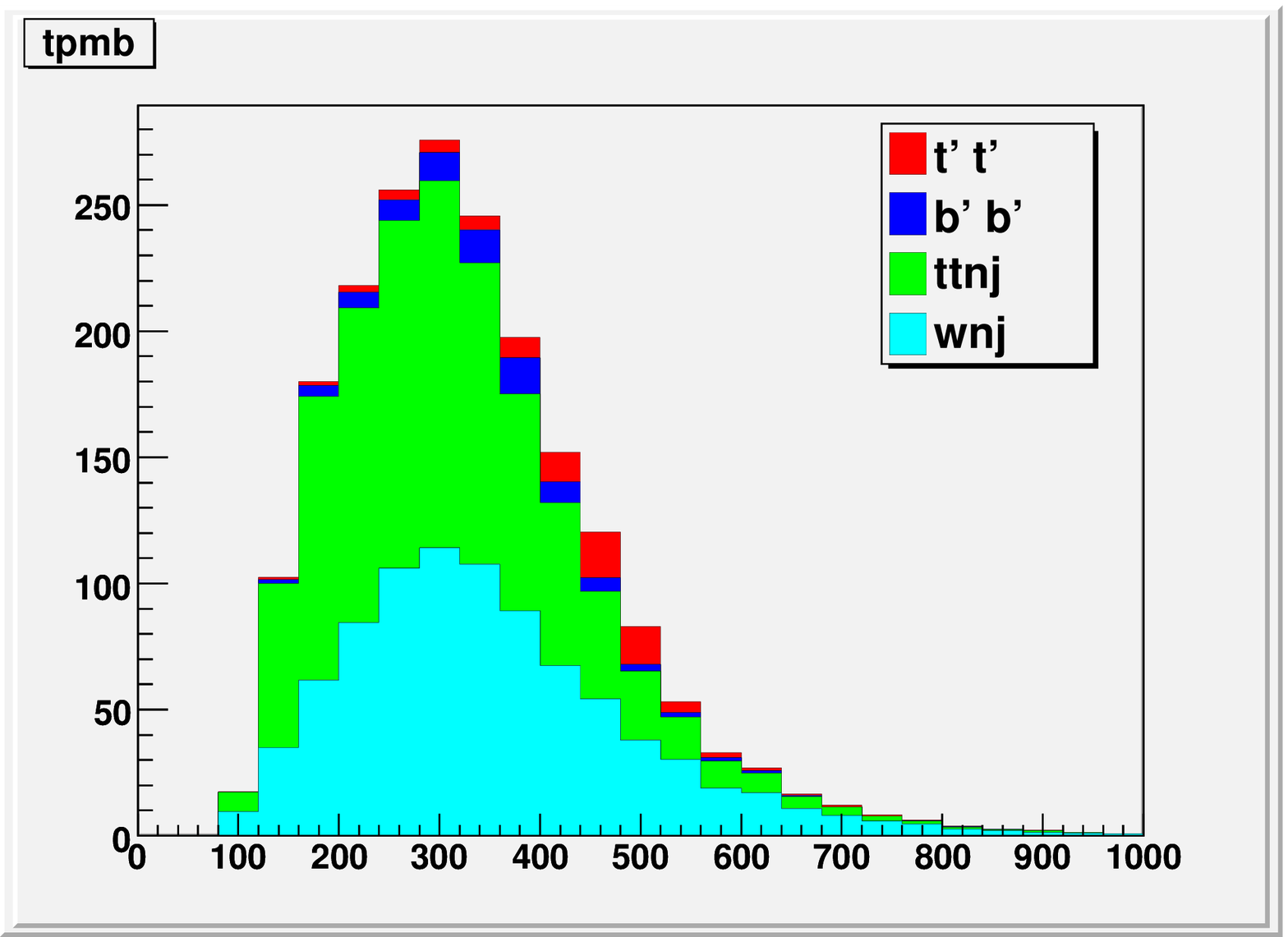}}
\vskip -0.05cm \hskip 0.4 cm \textbf{( a ) } \hskip 4.8 cm \textbf{( b )}
\caption{(a) The reconstructed $t'$ mass with the $\chi^2$ given in Eq. (1) is compared with the CDF reconstruction method. (b) The stacked reconstructed $t'$ mass is shown with the background included.}
\label{figmt}
\end{figure}

For event reconstruction we minimize the following $\chi^2$, where only the leading four jets in each event are used. The one or two jets with $b$-tags (or the leading two if there are more than two $b$ tags) are used to exclude those jets in the reconstruction of the hadronic $W$.
\bea
\chi^2 &=& \sum_{i=1}^2 \frac{|m_{W_i}-m_W^{PDG}|^2}{\sigma_W^2} + \frac{|P_T(W_h) - H_T/4|^2}{\sigma_{P_T}^2}  + \sum_{i=1}^2 \frac{|P_T(b_i) - H_T/4|^2}{\sigma_{P_T}^2} + \frac{|m_{t^\prime_1} - m_{t^\prime_2}|^2}{\sigma_{t^\prime}^2}
\label{chi5}
\eea
The $t^\prime$ is reconstructed quite well by taking $\sigma_{W}=15$, $\sigma_{P_T}=20$, and $\sigma_{t^\prime}=25 \gev$. The results are similar to the CDF reconstruction method based on scanning a reference mass $m^{ref}$ \cite{cdfnote}, as demonstrated in Fig. (\ref{figmt}a). The difference between the mass bump shapes of these two methods can be partially accounted by the fact that our method also takes into account the possibility that the hadronic $W$ is one jet. It is clear from Fig. (1b) that a simple mass bump analysis is not adequate to separate signal from background.

\begin{figure}[h]
\centerline{
\epsfxsize=16 cm \epsfysize=4 cm \epsfbox{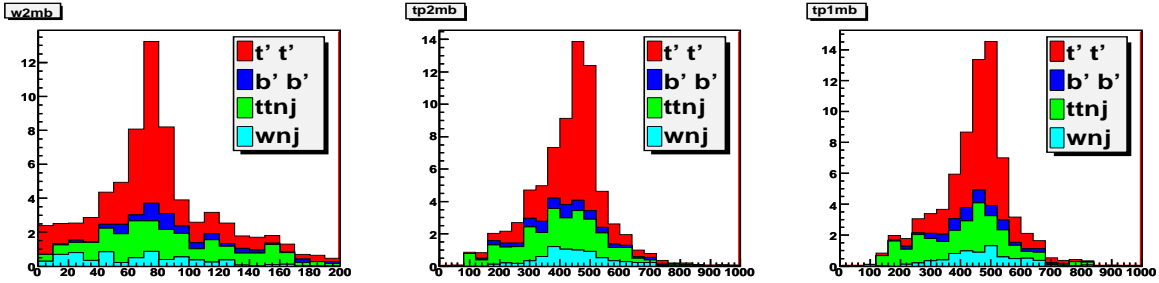}
}
\caption{Reconstructed masses after the MLP discriminant cut in Scenario 1 are stackedly shown.  Here the labels ``w2mb", ``tp2mb" and ``tp1mb" correspond to the reconstructed hadronic $W$, the reconstructed hadronic $t^\prime$ and the reconstructed semi-leptonic $t^\prime$. All are based on the $\chi^2$ defined in Eq. (\ref{chi5}).}
\label{figi}
\end{figure}
\begin{figure}[h]
\centerline{
\epsfxsize=16 cm \epsfysize=4 cm \epsfbox{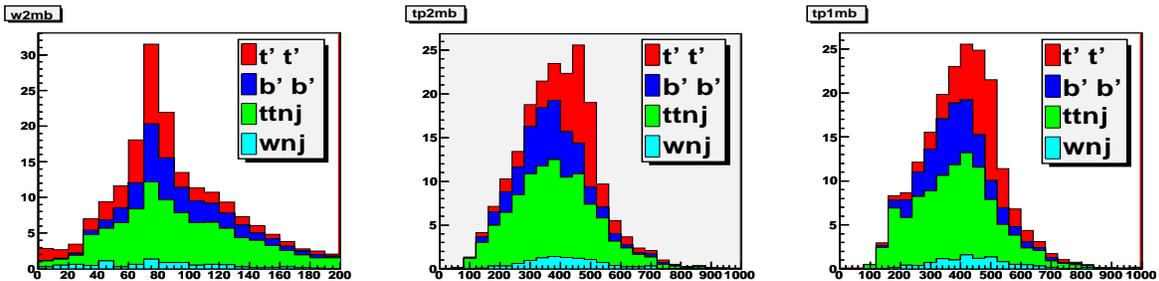}
}
\caption{Fig. (\ref{figi}) repeated for Scenario 3.  }
\label{figk}
\end{figure}

We use $b$ tagging efficiencies of 0.6, 0.1 and 0.01 for $b$, $c$ and light quarks respectively. Table \ref{table42} shows that $b$ tagging can effectively suppress the $W + nj$ background. The resulting significance is similar when the number of $b$ tags is either $n_b=1$ and $n_b=2$, and therefore we shall simply impose $n_b > 0$.

\begin{table}[th]
\begin{center}%
\begin{tabular}
[c]{|c|c|c|c|c|c|c|c|c|}\hline
& $b^\prime \bar{b^\prime}$ & $t^\prime \bar{t^\prime}$& $W + \textrm{jets}$&  $t {\bar t} + \textrm{jets}$ &$S/B$ &$S/\sqrt{S+B}$\\\hline
$n_b=0$ & $36.3$ & $20.1$ & $813.6$ & $433.0$ & $0.02$ & $0.55$\\\hline
$n_b=1$ & $ 48.8$ & $ 37.5$ & $61.8$ & $571.0$ & $0.07$& $1.4$\\ \hline
$n_b=2$ & $27.7$ & $28.0$ & $6.4$ &
$310.5$ &0.1& $1.5$\\ \hline
\end{tabular}
\end{center}
\caption{Number of events with tagged $b$ jet multiplicity samples ($1$ fb$^{-1}$) are demonstrated.}%
\label{table42}%
\end{table}

For the MLP neural network we choose 40 observables as input. The relatively large number of observables will more accurately  capture the phase space structure of both signal and background events, and our particular choice of observables has been optimized by trial and error. We consider three scenarios for how the $b'$ events are used in the training of the neural network. The results are presented in Table \ref{table4b11} where the degenerate $t'$ and $b'$ masses are taken to be either 500 or 600 GeV.
\begin{table}[h]
\begin{center}%
\begin{tabular}
[c]{|c|c|c|c|c|c|c|}\hline
& $b^\prime \bar{b^\prime}$ & $t^\prime \bar{t^\prime}$& $W + \textrm{jets} $ & $t {\bar t} + \textrm{jets}$ &$S/B$ &$S/\sqrt{S+B}$\\\hline\hline
Scenario 1 500 GeV  &$4.4$ & $34.9$ & $5.6$ & $15.4$ & $1.4$ &$4.5$\\\hline
Scenario 2 500 GeV  & $7.3$ & $38.1$ & $6.9$ &  $20.8$ &$1.6$ & $5.3$ \\\hline
Scenario 3 500 GeV  & $33.7$ & $41.8$ & $ 7.8$ &  $55.6$  & $1.2$ & $6.4$ \\\hline \hline
Scenario 1 600 GeV   &$2.4$ & $11.2$ & $4.0$ & $8.2$ & $0.8$ &$2.2$\\\hline
Scenario 2 600 GeV  & $3.0$ & $11.5$ & $3.9$ &  $9.3$ &$1.1$ & $2.8$ \\\hline
Scenario 3 600 GeV  & $9.8$ & $13.0$ & $5.5$ &  $25.9$ & $0.8$ & $3.1$ \\\hline \hline
Scenario 1 500 GeV  & $7.2$ & $36.2$ & $13.1$ & $31.7$ & $0.7$ &$3.8$ \\\hline
\end{tabular}
\end{center}
\caption{Summary of number of events (with $1$ fb$^{-1}$) from different scenarios for the treatment of the $b'$ events are shown. The last row shows  a sample result when using observables that do not rely on a full reconstruction. }%
\label{table4b11}%
\end{table}

\begin{figure}[t]
\centerline{
\epsfxsize=7 cm \epsfysize=5 cm \epsfbox{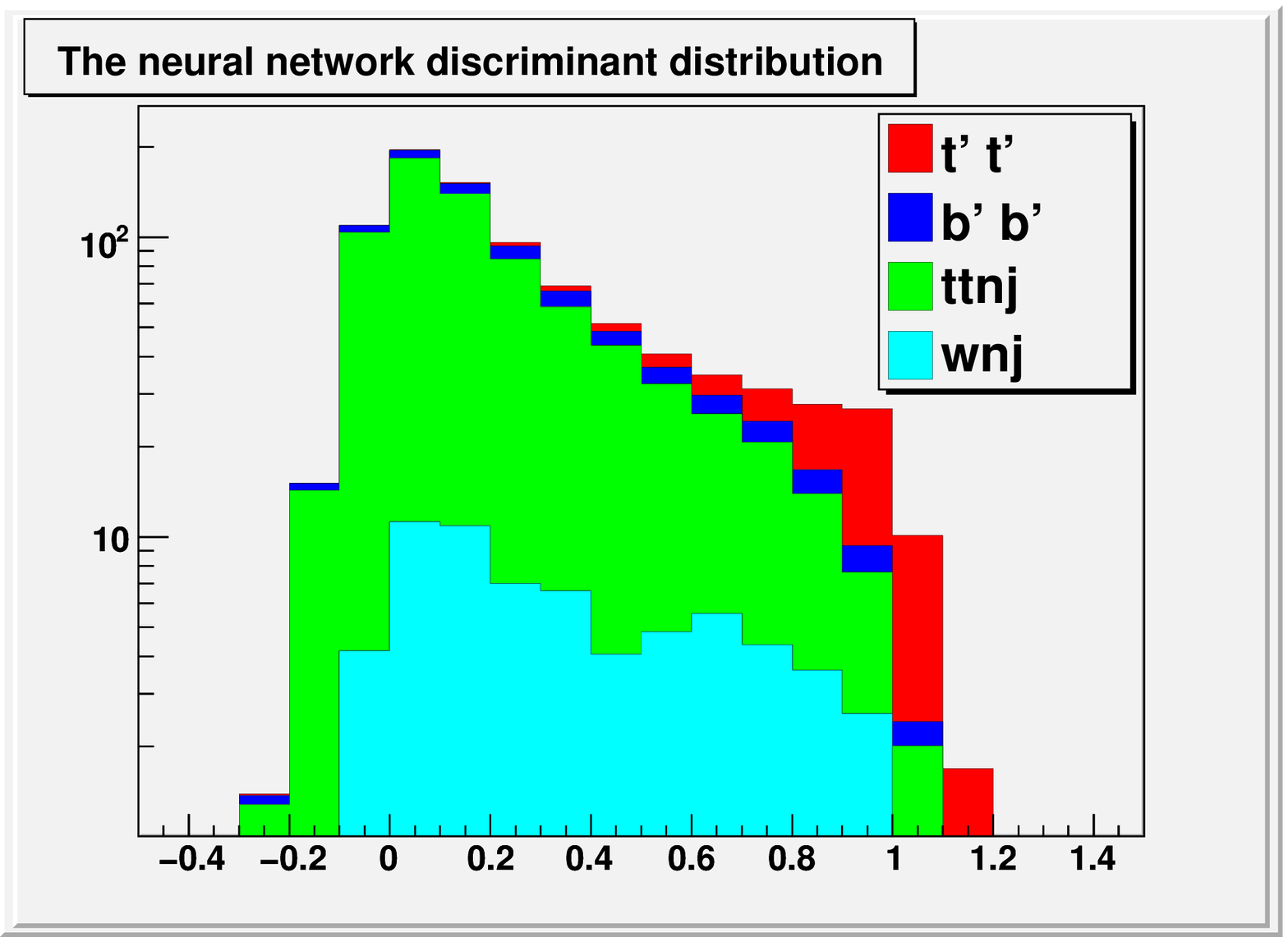}
\hspace*{0.2cm}
\epsfxsize=7 cm \epsfysize=5 cm \epsfbox{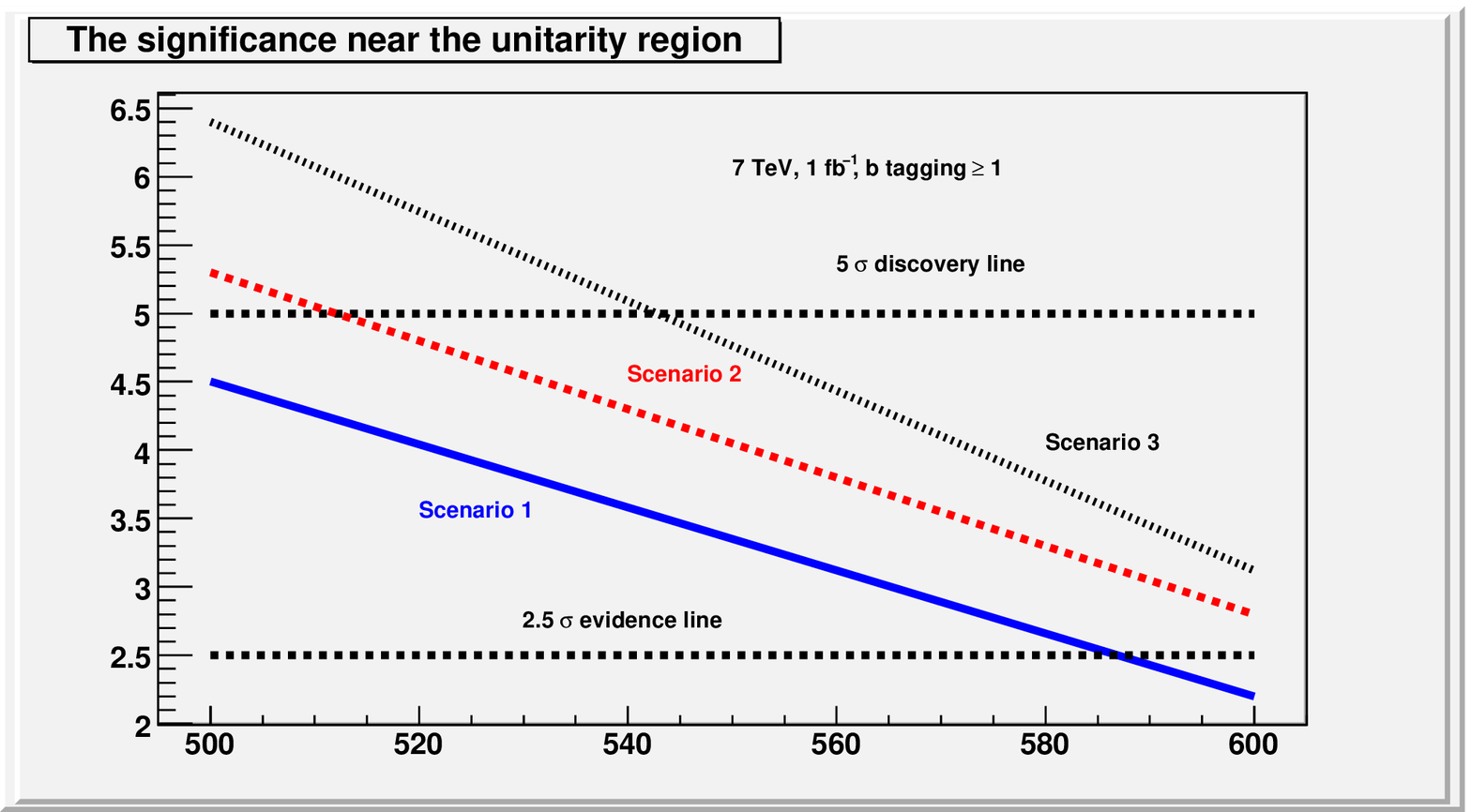}}
\vskip -0.05cm \hskip 0.4 cm \textbf{( a ) } \hskip 4.8 cm \textbf{( b )}
\caption{a) The stacked MLP discriminant distributions for signal and background for Scenario 1 are displayed. b) The sensitivities, $S/\sqrt{S+B}$, for the three scenarios at the early running of LHC with $7$ TeV are shown, where the x-axis is the mass of $t^\prime$ in GeV and the y-axis is the significance. }
\label{figf}
\end{figure}

1) In the first scenario the neural network is trained by treating the $b'$ events as background. This scenario helps to establish how well the signals from $t^\prime$ and $b^\prime$ can be distinguished. In Fig.~(\ref{figi}) we show mass reconstructions for the hadronic $W$, the hadronic $t^\prime$ and the semi-leptonic $t^\prime$ after applying the MLP discriminant cut. This figure demonstrates that both our reconstruction methods and the MLP method work quite well. The discriminant plot for this method is shown in Fig. (\ref{figf}a).

2) In the second scenario we ignore $b^\prime$ events in training. With respect to the neural network discriminant the $b^\prime$ events are distributed quite uniformly, and the events above the cut will contribute to the signal. The background rejection also improves a little and the result is a better $S/B$.

3) In the third scenario we train using both $t^\prime$ and $b^\prime$ events as signal. Note that the reconstruction procedure remains the same and is still geared towards reconstructing the $t^\prime$. This shows up as what appears to be a reduced $S/B$ in the three mass reconstructions of Fig. (\ref{figk}). In addition the location of the effective $t'$ mass peak shifts down, due to decay products of the $b'$ being missed. However the power of the neural network becomes apparent in this case, since it resorts to other kinematic observables to distinguish signal from background. Its success shows up in the nicely improved sensitivities in Table \ref{table4b11}.

For comparison we also show the performance that can be obtained without performing any reconstruction. In this case we only make use of the kinematic observables adopted in the top quark precision measurements \cite{Abazov:2007kg}. The result is shown in the last row of Table \ref{table4b11} and the comparison to the first row shows that observables associated with the reconstruction are somewhat helpful to achieve a better sensitivity.

From the results in Table III we display in Fig.~(\ref{figf}b) our final estimates for the sensitivity of the LHC to the $t^\prime$ for masses close to the unitarity bound. 

We note that there is an uncertainty on the overall normalization of the backgrounds due to our reliance on a Monte Carlo estimate and in the end a more data-driven approach to background subtraction may be adopted by experimentalists. But to gain some confidence in the Monte Carlo estimates we can test their stability by changing the parameter $p_{T\rm{min}}$ used by Alpgen in the generation of the dominant $t\overline{t}+\mbox{jets}$ background. We generate samples with jet multiplicities 0, 1, 2 and $\ge3$ for the two choices $p_{T\rm{min}}=100$ GeV (our choice above) and $p_{T\rm{min}}=20$ GeV (a more time consuming choice). The characteristics such as the $H_T$ distribution of the combined samples for the two cases are very similar as expected \cite{Holdom:2007ap}. In Table \ref{tablecomp} we present the number of events passing the discriminant cut of the neural network and we see that while the individual samples vary dramatically, the difference in the total number of events is small. This indicates that not only is Alpgen performing well, but there is no nontrivial dependence on $p_{T\rm{min}}$ arising through interaction between the event generation and the neural network.

\begin{table}[th]
\begin{center}%
\begin{tabular}
[c]{|c|c|c|c|c|c|c|}\hline
&$P_{T\rm{min}}$& $t{\bar t} +0j$ &  $t {\bar t} + j$ & $t {\bar t} + 2j$ & $t {\bar t} + (\ge3j)$ & $\textrm{total}$ \\\hline
Scenario 1 & 100 GeV & $4.2$ & $8.4$ & $2.6$ &  $0.2$ & $15.4$ \\ \hline
Scenario 1 & 20 GeV  & $1.4$ & $4.3$ & $4.2$ &  $4.6$ & $14.6$ \\ \hline \hline
Scenario 2 & 100 GeV & $5.7$ & $10.9$ & $4.0$ &  $0.2$ & $20.8$ \\ \hline
Scenario 2 & 20 GeV & $2.0$ & $4.8$ & $4.6$ &  $7.0$ & $18.4$ \\ \hline \hline
Scenario 3 & 100 GeV  & $11.5$ & $28.9$ & $15.1$ &  $0.2$ & $55.6$ \\ \hline
Scenario 3 & 20 GeV & $2.1$ & $8.2$ & $20.0$ &  $26.0$ & $56.4$ \\ \hline \hline
\end{tabular}
\end{center}
\caption{Number of events with two different $p_{T\rm{min}}$ values which passed the neural network discriminant cut are shown in three scenarios. }%
\label{tablecomp}%
\end{table}

Next we explore the future Tevatron bounds by assuming an integrated luminosity of $10$ fb$^{-1}$. The main results are shown in Table \ref{table50}, where the constraint on $n_b$ is released (due to the relatively poor performance of $b$ tagging at the Tevatron) while $n_j\geq 4$ is applied. Here we see an even bigger increase in the sensitivities in the progression through the three scenarios. Thus if the goal is to discover or rule out a fourth family one should allow both the $t'$ and the $b'$ to contribute to a signal. We see again that the neural network can actively pull out the combined signal from background even with a reconstruction method that is geared to the $t'$.

\begin{table}[th]
\begin{center}%
\begin{tabular}
[c]{|c|c|c|c|c|c|c|}\hline
& $b^\prime \bar{b^\prime}$ & $t^\prime \bar{t^\prime}$& $W + \textrm{jets}$ &  $t {\bar t} + \textrm{jets}$ & $S/B$ &$S/\sqrt{S+B}$\\\hline\hline
Scenario 1&$15.19$ & $30.52$ & $8.8$ & $41.6$ &  $0.46$ &$3.11$\\\hline
Scenario 2 & $22.10$ &$38.76$ & $13.5$ & $38.0$ & $1.18$ & $5.74$ \\\hline
Scenario 3 & $56.0$ &$41.2$ & $12.3$ & $94.3$ & $0.91$ & $6.81$ \\\hline
\end{tabular}
\end{center}
\caption{Tevatron results with $\sqrt{s}=1.98$ TeV and 10 fb$^{-1}$. Here $m_{t^\prime}=m_{b^\prime}=400$ GeV and $n_j \geq 4$.}%
\label{table50}%
\end{table}

In the current CDF $t'$ search \cite{cdfnote} the possible existence of a $b'$ is not considered. A $b'$ can generate events that could be interpreted as $t'$ events, especially given that a bin with 5 or more jets is kept. We have noted above that the $b'$ events can produce a distribution in the reconstructed mass $M_{rec}$ that is somewhat broader and lower than for the $t'$ events, assuming equal masses. The net effect could be to roughly double the cross section for the production of events in the high $M_{rec}$ and $H_T$ bins. This suggests that a reanalysis of the CDF data may be warranted, especially given the slight excess already seen in the high bins.

We have considered a search for $t^\prime$ type heavy quarks during the early running of the LHC with collision energy 7 TeV and integrated luminosity $1$ fb$^{-1}$ and found sensitivity to quark masses close to the unitarity upper bound, which is in the $500$ to $600$ GeV range. Any enhancement of the collision energy or luminosity would give good prospects for the discovery of a fourth family even at the high end of this mass range. In order to obtain more reliable estimates of the required luminosities a more detailed experimental analysis and full detector simulation is inevitably needed, which is beyond the scope of this work. 

We have noted that the sensitivity depends on how the $b'$ events are treated in the analysis. If the goal is to set limits on the existence of a sequential fourth family then it is advantageous to enhance the $b'$ contribution to the signal. A multivariate tool such as a neural network is an efficient way to accomplish this since it can account for the different characteristics of $t'$ and $b'$ events simultaneously. The CDF collaboration has presented the limits $m_{t'}<335$ GeV \cite{cdfnote} and $m_{b'}<385$ GeV \cite{cdfnote1} in separate analyses. We are suggesting that their sensitivity to the existence of a nearly degenerate quark doublet of a fourth family should be greater than these numbers indicate.

\end{document}